\documentstyle{lamuphys}
\makeatletter
\let\chapter\hid@chapter
\makeatother
\def\kms{~km s$^{-1}$~}
\begin{document}
\pagenumbering{arabic}
\titlerunning{Cosmological Questions for the VLT}
\title{Cosmological Questions for the European Southern Observatory
Very Large Telescope\footnote{
To appear in: 
ESO Workshop on {\sl The Early Universe with the VLT}, 
European Southern Observatory, Garching b. Muenchen, Germany 1-4 April, 1996
ed. J. Bergeron, (Springer: Heidelberg) (15 printed pages).
}}

\author{David\,Tytler}

\institute{University of California San Diego,
0111, La Jolla, CA 92093-0111, USA.}

\maketitle

\begin{abstract}
The next decade promises an observational revolution which will
change cosmology forever. The precise measurement of the angular
anisotropy of the cosmic microwave background should specify to a few
percent all of the parameters of the cosmological model which effect 
astrophysics. The growth of structure will then
be determined (but not yet observed) until gravitational collapse becomes 
highly non-linear and stars, galaxies and active galactic nuclei (AGN) form. 
Not all of the gas enters these collapsed objects, in part because of feed back:
stars eject gas, and both stars and
AGN emit ionizing radiation which heats the gas.  Instead, the 
universe enters a prolonged out-of-equilibrium state, which we are in today, 
when its appearance is determined by the balance between three things: 
the parameters of the cosmological model,
the efficiency with which matter enters various collapsed objects, 
the matter and energy released by those objects.
The competition between these collapse and feedback processes determines the 
appearance of the universe today, and their
study will replace the specification of the model as the long 
term focus of research in cosmology.

These processes are hard to model with basic physics because they are
complex and allow a rich variety of expression.
Instead observations will
determine when the first stars and quasars formed, and how and when galaxies 
assembled. These processes will be parameterized, and used to
improve computer models, which will be tested in our own and other
near-by galaxies. If we succeed, and reconcile the
numerous contradictions which characterize the subject today,
cosmology will become a mature subject, founded on the agreement between
detailed, inclusive and realistic models, which make precise predictions, 
and the wealth of new data which will come from a wide variety of 
observations, at all wavelengths.
This is an ambitious schedule, but nothing less is worthy of the
outstanding capabilities of the 8 -- 10 m telescopes, the 
next generation space telescope, the opportunities at millimeter
to sub-millimeter wavelengths and advanced computer modeling.
The ESO Very Large Telescope (VLT) should play a major role in this revolution.

\end{abstract}
\section{What is Cosmology?}
Cosmology, the study of the universe on the largest scales of space and time,
began with ancient creation myths which described the 
beginning of the universe and the Earth, the differences
between heaven and Earth.
In this century General Relativity provided a dynamical framework
for a variety of cosmological models, which allowed mathematical
discussions of the growth of perturbations.
Observations with the Mount Wilson reflector showed that galaxies
are the building blocks, and that the universe is expanding.
Experimental verification of the Big Bang came in the 1960's with the
detection of the evolution of radio galaxies and QSOs, the discovery of
the cosmic microwave background (CMB), and the prediction of the
abundances of the light elements from big bang nucleosynthesis (BBNS).
There followed much interest in the physics of the early universe, 
especially the growth of perturbations to form galaxies, and the
prediction of how different types of perturbations could be
distinguished in the anisotropy of the CMB 
and the motion and distribution of galaxies today.
In the 1980s the suggestion of inflation at the GUTs energy scale 
emphasized that the model was incomplete, and stimulated
intense interest in phenomena at the highest energies.
The subject is now extremely active, on many frontiers, and changing rapidly.

 From an observational perspective, the
major questions in the subject can be grouped under a few headings:

\smallskip\noindent
{\bf Parameters of the Model.} The standard big bang
model is now specified in terms physical parameters, which can be measured.
We believe that these parameters have an origin in the physics of extremely 
high energies, beyond those which are well understood, but perhaps 
they will be predictable with new physical theories.

\smallskip\noindent
{\bf Contents of the Universe.}
Which particles are important in the Universe?
Where are the baryons, and how many of them are missing in dark matter?
What are the main forms of non-baryonic matter?
What media fill the universe (intergalactic medium, background radiation 
field, gravitational waves)? What are the main objects 
  (large scale structures, clusters of galaxies, galaxies, active galactic nuclei, stars)?

\smallskip\noindent
{\bf Origins.} How, why and when did the particles, media and
objects form?

\smallskip\noindent
{\bf Evolution.}
How have they changed on cosmological time scales?

\smallskip\noindent
{\bf Processes.} What are the main
physical processes which affect origins and evolution?

\smallskip\noindent
{\bf Ecology.} How do objects interact, merge and feed back on
the media from which they form?

\section{Much Cosmology will be done Locally}
The frontiers of astronomy, including cosmology, have moved out in
distance (solar system, stars, our Galaxy, other galaxies, quasars, the
cosmic microwave background) and back in time (solar system formation, 
oldest stars and QSOs, CMB, primordial nuclear synthesis, baryosynthesis,
inflation, ...)
but in addition to direct observations of the most distant objects, much of the
critical cosmological information will continue to come from the detailed
study of local objects, and there could be 
laboratory detections of non-baryonic dark matter.

\subsubsection{Oldest Stars.} The ages, orbits and chemical compositions of
the oldest halo stars are the best clues to the formation of our Galaxy,
and are critical to our understanding of primordial element abundances,
and chemical evolution.

\subsubsection{Structure of Galaxies.} Our Galaxy and other near by galaxies 
are the best places to understand the distribution of dark matter,
the relative and absolute ages of stellar populations, and galactic
chemical evolution (how much gas went into stars? what types of stars?
what types of remnants did they leave?).
 
\subsubsection{Local Objects and Structures.}
Detailed understanding of the different types of galaxies, of active
galactic nuclei (AGN),
clusters of galaxies,  and of large scale structure, streaming motions and 
biases in the distribution of galaxy light relative to mass
will continue to come from studies of the nearest examples.

\subsubsection{Understanding key Cosmological Processes.} Again many, but not
all of the key processes can be observed locally:
star formation, galaxy mergers, supernovae.

\section{Determination of the Cosmological Model}
The apparent detection of the first doppler peak in the CMB 
(White, 1996) suggests that
this radiation was last scattered at the recombination epoch, and
still contains information from that time. In less than 10 years
ground based interferometers, balloon born detectors, and especially the 
COBRAS/SAMBA and MAP
satellites should together measure the angular power spectrum 
of the CMB on scales $l<1000$ (1000 independent pieces of cosmological
information) to within the limits of cosmic 
variance and the (benign) Galactic and extragalactic sources.
This data should specify the values of all the main cosmological 
parameters to high precision: COBRAS/SAMBA \cite{cob} now promises 
$H_o$, $\Omega_o$, and $\Lambda$ to 1\% and
$\Omega_b$, $Q_{rms}$, and $n_s$ to few percent (Jungman et al. 1996;
Hu, Bunn \& Sugiyama 1995).

COBRAS/SAMBA will also measure the spectrum distortion parameter $y$ in
$>10^4$ clusters of galaxies, giving estimated of $H_o$ from X-ray data,
the cosmological evolution of clusters, cluster bulk velocities 
to 50~km~s$^{-1}$ out to $z=1$, large catalogues of IR galaxies, radio 
galaxies, AGN and counts of normal galaxies.

\subsection{The Microwave Revelation}
Precise measurement of the cosmological parameters will revolutionize all of 
cosmology. Models will make precise predictions, and
observations will directly yield absolute physical quantities
(redshift will yield distance, and look back time; apparent magnitude will
yield luminosity; angles and redshifts will give linear separations).
The classical cosmological tests from the 1960's, designed to measure the
parameters of the model, will be more relevant as consistency checks and
especially measures of evolution (Gunn 1977).
Indeed, we predict that the model will be so well determined that 
evolution, especially of
complex structures like galaxies, AGN and stellar populations,
will replace the determination of the model as the main goal of 
cosmology. And we should expect surprises,
perhaps coming from todays list of problems:
the age problem, $\Lambda$, the types of dark matter, $\Omega \ne 1$,
non-gaussian fields (strings), strong gravity waves, early re-ionization of
the intergalactic medium (IGM), isocurvature perturbations, ... 

\section{Checking the Model}
Before the main CMB data arrive (2002), and after, if they are complex to 
interpret, we will check the parameters of the model.

\smallskip\noindent
{\bf $\Lambda$} can be determined from the number and distribution in
$z$ of gravitationally lensed QSOs (Kochanek 1996).
The Sloan digital sky survey should provide the 100 - 1000 lensed QSOs.

\smallskip\noindent
{\bf $\Omega$} can be determined from supernovae out to 
$z \simeq 1$ (\cite{goo}; \cite{lie}), and from lenses.
 
\smallskip\noindent
{\bf $H_0$} can be determined from a few simple lens systems and supernovae.

\subsection{Time Scales}
The CMB should give the age of the universe $t_0$ and $t(z)$: 
our first  well defined cosmological time scale.
It will then become critical that we know the ages of stars as
precisely as possible, because they date the key events in galaxy formation.
Globular cluster ages, now known to 10 percent, should be known to a 
few percent.
One percent seems too hard at this time, because each of the following
errors gives a 1 percent age error on its own:
$\delta$(distance modulus) $=0.01$  mag,
$\delta$(Helium abundance)  $=0.01$,
$\delta$([Fe/H])            $=0.03 $ dex,
$\delta$([alpha elements/Fe])$=0.03 $ dex,
$\delta$(E(B-V))            $=0.003 $ mag.
The first two may be reachable, but the last three items are beyond
hope (Renzini, private communication, and \cite{ren}).

Very complete luminosity functions, from the turnoff to lower giants,
and improved data are needed.
We can also use the VLTI to give 10 $\mu$arcsec parallax measurements for 
globular clusters, provided we find or obtain reference stars of
known distance or position within the isoplanatic angle. 
This gives a 1 percent distances error at 1~kpc.
Rough checks can be made with eclipsing binaries at or below the turnoff.

Improved ages will also be obtained from radioactive chronometers, 
especially Thorium in halo stars. Here we need to find tens of stars
which have low metal abundances (to avoid blending with metal lines)
and enhanced $r$-process elements. Only one is known today.
It is not know if the VLT will uncover new cosmological clocks.

\subsection{Thorough Tests of Big Bang Nucleosynthesis}
The accurate measurement of the abundance of the light elements is
an ideal project. The ratios of the abundances of different nuclei
test the predictions of BBNS, while the variations of these abundances 
at later times are excellent tests of the predictive ability of our 
understanding of stellar interiors (especially mixing), and stellar 
and Galactic chemical evolution.  These observations are well suited 
to the VLT, and there are many outstanding problems,
which will not be solved for many years.

There are two observational approaches. First, and ideally we would measure abundances
in primordial gas. QSO absorption systems often have low abundances
[C/H] $<-2$, which are low enough, but there are no cases with 
[C/H] $<< -3$, because such low abundances do not occur where
total column densities are high enough to allow such low limits.
Second, we can seek to understand how abundances change over time, using
the ratios of various nuclei measured in different places.

\subsubsection{Deuterium:} 
This nucleus is made in BBNS (only) and destroyed in stars.
The 1215\AA ~line, 82 \kms to the blue of the H Lyman-$\alpha$ line
has been seen in at least two QSO absorption systems which have low [C/H]
(Tytler, Fan \& Burles 1996; Burles \& Tytler 1996). If the other
possible detections (Songalia et al. 1995; but see \cite{tyt-atami})
are also real, then D/H may vary spatially.
We believe that it can be seen in about 3 percent of QSOs at $z_{em}=3$, so a
search of 3000 QSOs would give 90 D/H measurements.
If primordial D/H is high, there is a lot of destruction, which
is not understood. We would see how D/H correlates with metal abundance,
and measure the dispersion at a given abundance. The change in
D/H as a function of metal abundance depends on the masses of the stars:
high mass stars make more metals for each H atom returned to the 
interstellar medium (ISM).
However if the first 10 high quality measurements show that primordial 
D/H is low, and 
homogeneous, then we would stop the project at that time.

\subsubsection{Helium-3:} This nucleus is made in BBNS and low mass stars, and
destroyed in high mass stars. But this traditional description does not
account for the basic features of the data, and must be wrong.
The abundance in H~II regions varies from 1 to $8 \times 10^{-5}$. 
Low mass stars are expected to make and release a lot of $^3$He, but abundances
in H~II regions are an order of magnitude less than expected
(Galli et al. 1995).
It has been suggested that H~II regions are not representative of the ISM
because they contain winds from high mass stars which destroy $^3$He
(Olive et al. 1995).

\subsubsection{Helium-4:} Recent measurements of the abundance of this nucleus 
(Thuan, Izotov \& Lipovetsky 1996) do not 
agree with either the low or high values for deuterium seen in QSOs. 
Something is wrong, perhaps large systematic errors which should be corrected.

\subsubsection{Lithium-6:} This nucleus is created by cosmic ray spallation on
oxygen (and C, N) 
in the interstellar medium, and perhaps also in stellar flares.
It is extremely hard to measure, since its lines are only 7 \kms to the
red of those of the 6708\AA\ doublet of $^7$Li, and signal to noise of
hundreds is required, but this is an excellent project for the VLT.
It has been seen in only two stars. Measurement in others with different
metallicities and mass would tell us more about mixing since this
isotope is more readily depleted than $^7$Li.
It has been suggested (Copi, Schramm \& Turner 1996) that $^6$Li is the
``strongest" argument against depletion of $^7$Li, since $^6$Li is
more easily destroyed than $^7$Li inside stars, but this
argument is flawed. There may be other ways of making $^6$Li in stars,
and there may be ways of depleting $^7$Li without removing all of $^6$Li,
such as stellar winds (\cite{vau}), or gas with no Li which is
mixed to the surface.

\subsubsection{Lithium-7:} Multiple creation sites are considered: BBNS, 
cosmic ray spallation, AGB stars, supernovae, novae. 
Population I stars have more than we expect from BBNS, and
ten times more than low metallicity population II stars. 
It is destroyed inside stars, for example in 
low temperature population II stars where convection extends deeper and brings
depleted gas to the surface.
Warm halo stars with [Fe/H] $< -1$ show approximately constant
$^7$Li, the Spite plateau, which some cosmologists would like to treat as the
primordial abundance. But there are variations on the plateau, similar
stars have different Li/H, and there may be a correlation of Li/H with T$_e$
and possible [Fe/H] (\cite{rya}, \cite{mol}).
We must expect some depletion of Li because it
either gravitationally sinks into a star if the atmosphere is non-turbulent,
or it is convected down and destroyed inside
(\cite{vau}). The question of how much
depletion, and hence the determination of the BBNS abundance,
will be answered with more data.

\section{Astrophysical Input to the Model}
Many important parameters, especially those determined by complex 
astrophysical processes, will not come from the CMB measurements, and 
must be determined by other observations.

First, we need to determine the distribution of baryonic dark matter,
and the nature of the non-baryonic dark matter. From astrophysics alone
we get phenomological descriptions of the various ways in which the dark 
matter acts in cosmological situations. If different 
particles behave in the same way in all astrophysical situations, then
we will not distinguish between them, but we can pass these phenomological
descriptions on to particle physics.

Second, we need an observational description of the various
important non-linear processes: large scale structure formation, galaxy 
formation, star formation and AGN formation. We also need to map out 
non-linear evolution and the relevant feed backs.

Third, we need to measure the radiation and other backgrounds, and determine
the sources of radiation, and the extinction in dust (intergalactic and
inside galaxies and QSOs).

VLT observations can be used to test and guide computer
simulations, for example, by parameterizing star formation 
(initial mass function, efficiency, burst size and distribution) in terms of
variables ($\rho$, T, background ionizing spectrum $I_{\mu}$, abundances,
magnetic fields) which should be followed in simulations.

\section{Dark Matter}
First we must determine how many types of dark matter are dynamically
important on various scales.
Candidates include hot and cold particles, hot diffuse baryons 
(hard to detect because the H is ionized and the density is too
low for substantial X-ray emission), and baryons in collapsed objects
(stellar remnants, MACHOs, black holes).
We should attempt to measure the ratios of the densities of these
dark components, together with luminous matter, and we must work out
how the ratios vary with environment.

Radial velocities, proper motions and distances for selected stars,
together with gravitational microlensing will trace dark matter in our Galaxy 
and a few near by galaxies, including local dwarfs and M31, where
pixel lensing should be decisive.  Weak lensing 
is powerful for clusters of galaxies, while proper motions of galaxies out 
to Virgo (VLT Interferometer; \cite{tyt}) 
and improved galaxy distances will explore 
larger scales.  The lensing of QSOs is one way to find massive dark galaxies.

\section{Formation of Dense Objects}
Understanding the formation of dense objects (galaxies, AGN, stars, 
Pop~III objects) is central to cosmology after the cosmological model is 
determined. These objects dominate evolution of the observable universe.
They release energy which heats the ISM and intergalactic
medium (IGM), limiting their 
formation. They are the source of metals, which effect the cooling
of gas, and they release the ionizing radiation, which controls the
opacity of the ISM and IGM.
These complex feed back processes produce highly inhomogeneous media, with 
orders of magnitude variations in density and temperature.

At early times, galaxies, clusters, QSOs all look different.
We will need to find ways to place objects at different $z$ on their
correct evolutionary tracks.

\subsection{Formation of Galaxies}
Galaxy formation extends over a large range of redshift, and is
not complete for at least 2~Gyr because of large dynamical time scales
in the outer halos ($>100$~kpc).
The epoch of formation depends on the spectrum of primordial perturbations.
For CDM models most mass in galaxies was accreted at low $z$,
and virialized systems at $z=4$ were of sub-galactic mass.
This will be tested, for example
with the splitting of QSO lenses
(large samples are needed to get many lenses at high $z$), which measure 
$\sigma_v$, and with the velocity extent of Quasar absorption lines
(\cite{wol}).

Peebles (1988) has noted that different aspects of galaxy formation may
happen at different times:

\smallskip
$z_b$: the assembly of 0.5 of the mass in bulges/halos today,

\smallskip
$z_d$: the assembly of 0.5 of the mass in disks today,

\smallskip
$z_s$: the time of formation of 0.5 of the mass in long lived stars seen
today.

\noindent
The last can be measured with the images and spectra of high $z$ galaxies.
How do these epochs depend on galaxy morphology, mass, and environment 
(voids, clusters)?
How much mass do galaxies eject in hot winds? What is the metal contents
of those winds? Were the first pre-galactic objects Pop III stars?

\subsection{First Stars}
When do first stars form, especially when in
relation to the formation of QSOs and galaxies?
The recent discovery that the spectra of common galaxies at $z=3$ can
be obtained with the Keck telescope prompts us to prepare for
major new investigations of the
properties of young galaxies, perhaps reaching back to the earliest stars.
Steidel and colleagues (\cite{mac}; \cite{gia})
find that galaxies with
a surface density of 0.4 per arcmin$^{-2}$, and $R<25$ have half the
density of galaxies with $L>L^*$ today.  Star formation rates are about
$4-25~ h_{50}^{-2} M_0 yr^{-1}$ ($q_0=0.5$), and the
majority have 0.7 arcsec ($8.5 h_{50}^{-1}$kpc) cores and 
half light radii of $0.2-0.3$ arcsec, similar to galactic bulges. A minority
have exponential light distributions.
There are immediate opportunities to make larger samples, explore
higher and lower (HST for UV images) redshifts, and
lower luminosities, to derive the luminosity function, masses, morphology, 
dust contents, clustering and large scale structure at these times.
These techniques might find ``proto-clusters''.

A related question is the origin of the carbon seen in
the high column density Ly-$\alpha$ forest. Was this carbon made by the first
stars? Did those stars have a wider distribution sufficient 
to contaminate much of the volume of the universe, or did the carbon come from
from stars which formed in the structures which we see in absorption?
Alternatively, the carbon could be ejected
from pre-galactic units which later merged to form galaxies.

Were the first stars all of high mass, so that none exist today, and
are there other types of objects at high $z$, objects not seen today?

\subsection{First Active Galactic Nuclei}
Some QSOs at $z>4$ are of such high luminosities that they have $R=19$.
Objects of this luminosity could be seen at much higher $z$ if they
exist and there is no obscuration. 

Are QSOs the first collapsed objects?
When do they form in relation to the stars in galaxies?
If they form in the rarest of high density peaks, then we expect to see
galaxies clustered near by.

What determines the rate of increase of the central mass 
(\cite{hae}; \cite{ume})?
How efficiently does gas loose angular momentum and sink to the
center of a galaxy?
What determines luminosity, and are the first QSOs like those seen at $z<4$?

How and why do QSOs evolve? How long are QSOs luminous?
What determines if there is radio emission?
How common are low luminosity AGN at high $z$?
What fraction of galaxies contain massive black holes as a function of
epoch? About 46\% of all nearby galaxies having B$_T <$ 12.5 mag have 
AGNs, if one includes LINERs in the AGN (\cite{ho}, \cite{fil}).
How does black hole mass relate to galaxy morphology and
mass? What determines which galaxies are QSOs today --
the fueling of the black holes by galaxy interactions?

QSOs at $z < 7$ may be found with wide field IR
surveys (\cite{mcm}). We need a wide field telescope
with IR detectors across the focal plane, and 4-m telescopes to
identify those objects with the colors of QSO candidates.

Why do QSOs cluster around local galaxies? Lensing apparently does not
explain this.

\section{Galaxy Evolution}
I shall not discuss galaxy evolution, which is covered by several
contributions to this volume, except for the following.

Population synthesis from integrated spectra
can be used to determine the distribution of relative ages and abundances
in high redshift galaxies. In local galaxies we can use HR diagrams
(\cite{ort}).

Clusters of galaxies are laboratories for galaxy and star formation.
Like other types of clusters, they are special targets,
because many galaxies are observed at once, and all are at similar
distances, although probably not of the same age.
We need to find samples of clusters, and forming clusters at
$z>1$ to determine the
star formation rate, galaxy morphology, cluster mass, and mass to light ratios
all as function of $z$ and environment.
Samples at different $z$ may be matched by comparing comoving densities.

Data now hint that the regions with the most active star formation may 
change in time:
$z \simeq 3$ halos/bulges,
$z \simeq 2$ ?,
$z \simeq 1$ irregular blue galaxies,
$z \simeq 0$ spiral disks.
VLT images and spectra should correct this speculation,
and show if there is a dependence on mass and environment.

\section{Universal Chemical Evolution}
Quasar absorption lines should provide
the cosmological averages densities of various common elements:
$\Omega_{element}(z)$. We can also measure element abundance
ratios (e.g. $\Omega_O(z)/\Omega_C(z)$) which indicate the origins of the
elements (\cite{pet}; \cite{lu}).
Dispersion in ratios at a given time relates to the amount of mixing.

Most common elements have been seen in Quasar absorption lines:
H,  D,  He, C, N, O, Na, Mg, Al, Si, S, Ca, Ti, Cr, Mn, Fe, Zn and Ni.
In the most favorable cases, we may detect some of the following (\cite{ver}):
Li, Be, B, Cl, Co, Cu, Sc, V, Ga, Ge, Cd, Sn, Ba and Pb. We need
damped Ly$\alpha $ systems with 
log~N(H~I) $=21.5$, 0.1 solar abundances, and signal to noise 1000
spectra. At lower abundances, which are more interesting, the lines
are too weak to see.

Much more detailed studies of element ratios, and their origins
can be made on halo stars.
The distribution of halo stars as a function of metallicity tells us
about the first stars, while their element abundances tell about the first
supernovae (\cite{mcw}). Perhaps the elements in some halo stars
were made by individual supernovae, before the gas from many stars 
had mixed to yield average abundances.

\section{Intergalactic Medium}
Here, as in other areas, we have yet to determine the basic
parameters. The mean density of the IGM could comprise a few percent of
all baryons, or most baryons, with high values preferred by
the ionization of the Ly$\alpha$ forest (\cite{peti}, \cite{bi},
Rauch et al. in preparation).
The density determination is difficult because the IGM is highly
ionized, and we see the small amount of gas which is neutral.
When was the IGM re-ionized?
We expect different epochs for H and He, depending on the spectrum
of the ionizing energy.
The reionization, and galaxy formation will lead to variations in
density and temperature.
We can also hope to determine the origin -- galaxies, QSOs -- of the
background ionizing radiation using the proximity effect on
QSO absorption lines. We may be able to measure how the intensity and
spectrum of this radiation varies spatially and in time.

We expect galaxies and pre-galactic clumps to eject some, or even
a lot of metal enriched gas.
The gas in clusters of galaxies is very metal rich, perhaps in
part because galaxies everywhere eject gas, not just those
in clusters, where the gas is hot enough to see.
How do abundances vary in the IGM, with epoch and
especially with respect to the masses and types of the nearest galaxies,
and past environment (what types of galaxies in this 
neighborhood in the past?).

One outstanding question is the determination of the amount of dust
opacity on cosmological distances (\cite{fall}).
Are there dust obscured QSOs (\cite{web})? 
There are two ways to approach this question. First we can estimate the
dust contents of $z=3$ star forming galaxies
(Steidel's galaxies are biased against dust).
Second, we can use element ratios to measure dust in
damped Ly$\alpha$ systems.

\section{Large Scale Structure}
We can distinguish three epochs by the mode of observation: 
$z \simeq 1000$ can be observed in the CMB,
$z <1$ at $r \simeq 21$ can be observed with existing instruments like the 
Two Degree Field of the Anglo-Australian Telescope and the Sloan Digital 
Sky Survey.  The VLT should then plan to work on the intermediate redshifts, at
$r >> 21$.
We can also hope to use QSO absorption, QSOs and clusters of galaxies
to supplement galaxies in the determination of structure at high $z$.

\section{The Full Cast of Players}
Thirteen mirrors with apertures $>6.5$~m, at good sites, and with new 
instruments will be operating in a few years. Some will be specialized.

The Sloan Digital Sky Survey
will cover 10,000 sq deg. with a 2.5-m telescope, giving the
positions and images of $10^8$ objects to $R=23$ ($5\sigma$), the
spectra of $10^6$ galaxies to $B=19$, and $<10^5$ QSOs to $B=20$.

The HST may continue to operate in no-repair mode past 2005 with much 
improved instruments, including NICMOS and STIS (1997) and the Advanced Camera
for Surveys (1999). 

The Next Generation Space Telescope, an 8-m optimized for 1-5 $\mu$ with a 
$2007$ launch, should have a dramatic effect on astronomy.
It should excel at its goals: the study of the first starlight, assembling 
pieces of galaxies, changes in morphology, interactions, ... indeed, most 
of the central cosmological projects for the ground based telescopes.
It will have vastly superior angular resolution, and sensitivity in
the IR, and for many projects it will easily out perform
all ground based optical/IR telescopes.
But in the decade before it is launched, we note that;
{\it ~~~~If it can be done from the ground, it will be done from the 
ground before a space mission is launched,} (\cite{bei}).

\smallskip
New millimeter and sub-millimeter arrays (\cite{sha}) should also have a dramatic 
effect on cosmology since 
the peak of dust emission redshifts closer to those
wavelengths, which allows the detection of
thermal dust emission at $z<20$ for $L>10^{11}L_o$,
far beyond what we expect to reach in the optical and infrared.
The Millimeter Array could detect CO emission from ordinary galaxies to 
$z=1$, and from ultraluminous galaxies to $z=3$.

Computer simulations will also improve enormously once we know the
cosmological model, and because of continued improvements in
computing power, perhaps by a million fold by 2002 (\cite{pw}). 
They will give much more precise predictions of what to expect.

\section{Strengths of ESO and its VLT}
The following strengths should be used to target particular projects in 
cosmology.

\smallskip\noindent{\bf $\bullet$} The VLT is the largest telescope.

\smallskip\noindent{\bf $\bullet$} It has the most varied complement of focal 
instruments.

\smallskip\noindent{\bf $\bullet$} The Southern site is best for the 
Galactic center, the bulge, the Magellanic Clouds, and for gravitational 
microlensing.

\smallskip\noindent{\bf $\bullet$} The community of astronomers is large.
If there is interest in coordination and trying out new ways to conduct 
research,
this could be turned into a major advantage, as data rates rise, and the
pace of research in astronomy as a whole increases, as it must with all
these new facilities.

\smallskip\noindent{\bf $\bullet$} The ESO community has lead the way with 
adaptive optics and has the most experience in this area.

\smallskip\noindent{\bf $\bullet$} The VLT Interferometer, with laser 
guide stars and AO on all four unit telescopes, and $\geq 3$
outrigger telescopes would be the best IR/near-IR interferometer.

\section{The Urgent Need to Find Targets}
Keck is running out of ideal targets for some projects!
It is absolutely essential to get new large samples,
including:

\smallskip\noindent {\bf $\bullet$} High $z$ clusters of galaxies.

\smallskip\noindent {\bf $\bullet$} 
$10^4$ QSOs to find damped Ly$\alpha $ absorption,
and measure the rise of metal abundance with epoch. 

\smallskip\noindent {\bf $\bullet$} 
3000 QSOs at $z \simeq 3$ to make 90 high quality measures of D/H.

\smallskip\noindent {\bf $\bullet$} A sample of $10^5$ QSOs would yield about
1500 lens candidates and 400 lenses.

\smallskip\noindent {\bf $\bullet$} A sample of 50 metal poor stars with
[Fe/H] $<-4$ should come from spectra of 5000 stars with 
[Fe/H] $<-2$ (\cite{beer}).

\smallskip\noindent {\bf $\bullet$} 
A sample of 100 -- 1000 QSOs at $z>5$ might require spectra of $10^5$ candidates
(McMahon, private communication, and this volume).

\smallskip\noindent
A variety of telescopes could be fully occupied finding targets.

\smallskip\noindent {\bf $\dagger$} 
2-m telescopes: $z$ for $10^4$ - $10^6$ QSOs, and rough abundances for 
5000 halo stars.

\smallskip\noindent {\bf $\dagger$} 
4-m telescopes:

\smallskip
200 nights for 3000 QSO spectra to find candidates for D/H measurement.

\smallskip
200 nights for $10^4$ spectra to find damped Ly$\alpha $ systems.

\smallskip
3000 nights to check 1500 lens candidates.

\smallskip
1000 nights for $10^5$ $z>5$ QSO candidates.

\smallskip
Light curves of SN~Ia at $z=1$ with $I=25$ at max. 
\smallskip

\smallskip\noindent {\bf $\dagger$} 
Wide field telescopes: find QSOs, find lensed QSOs, find low abundance halo 
stars, and find and follow up microlensing events.

We note that it would be natural for ESO to lead in microlensing
because Paranal is the best site for bulge observations
and the VLTI will determine the distances and masses of individual lens
objects (\cite{tyt}, \cite{mer}).

\section{Deep Wide Field Imaging Surveys}
We should begin new deep wide field surveys which
will open up new areas for research.
Examples from the past include the Shane-Wirtanen Lick galaxy counts,
which lead to an understanding of the distribution of galaxies
(\cite{pee2}),
the Abell cluster catalogue, and the 3C and related radio catalogues which 
showed that the universe is evolving.
These new surveys could include deep images, 
like the HST Deep Field, and spectra to produce deeper versions of the
Sloan Survey. Others should be specialized, and cover large areas to find
high $z$ clusters, QSOs, lensed QSOs, and supernovae.

It might be most effective to devote one 8-m to such surveys.
Many of the best projects are very large, and will require team work.
We should practice all stages of these large projects: the planing, 
allocation of resources, distribution of work, coordination of
analysis and the sharing of rewards. 
There will remain numerous opportunity for small projects because 
both the objects and the processes in cosmology are complex.

\section{Cosmology will become Richer}
In past centuries and decades the frontiers of cosmology have moved
out in distance, and back in time.
We predict that this trend will end, once we have
detected the earliest objects, perhaps in the next decade.
Instead the focus will move to the detailed understanding of 
know objects and their evolution. The subject will become richer,
more demanding, and more mature. 

\smallskip
{\bf COMPLEX OBJECTS.} Like geology and biology, and unlike particle 
physics, the objects in cosmology are complex.
Although we speculate that most types of objects have been discovered,
there is great variety within each type of 
(neutron star, white dwarf, brown dwarf, giants,
supernovae, ...) and this variety can help us understand.

\smallskip
{\bf COMPLEX PROCESSES.} The processes in astrophysics are themselves
complex, whether in dynamics, nuclear physics or gas dynamics.
Consider the variety of effects which occur in dynamical systems:
tides, tidal locking, orbit evolution, precession, stability, relaxation, 
mass segregation, evaporation, binary formation, mergers, accretion, 
fragmentation, ...

\smallskip
{\bf COMPLEX SYSTEMS.} The processes which connect different types of objects
make complex systems. The ISM is a rich example, with
mass sources:-- in fall, stellar winds, stellar ejecta;
mass loss:-- galactic fountains \& winds, mass stripping, star formation;
energy sources:-- cloud-cloud collisions, supernovae shocks, star light, cosmic
rays;
variations in phase:-- 3 stable phases of temperature and density.
processes:-- ionization, molecular formation, dust formation, 
radiative transfer, opacity. 
In contrast, our understanding of the IGM is primitive.

\smallskip
{\bf EVOLUTION.}
All objects had an origin, but there can be multiple origins for
objects which are later identical: e.g. star formation.
Objects and processes can be far from equilibrium, leading to
increased variety.
Most objects die, and there are multiple possible deaths:
e.g. planets merge, fragment, fall into stars,  and are ejected from systems.
Many processes are irreversible and destroy information about the
past. But even when the large scale picture does not reveal origins,
the details, such as element and isotope ratios, orbits and
spin axes may.
For these reasons of complexity, cosmology has a long and rewarding
future.

\smallskip
I should like to thank 
Chas Beichman,
Arlin Crotts,
Alex Filippenko,
Chris Kochanek,
Richard McMahon,
Alvio Renzini,
Michael Rich,
Peter Schneider,
Richard Simon,
Ed Turner,
Art Wolfe
and especially Jacqueline Bergeron for the invitation to consider this
topic, and for hosting this forward looking meeting.

%
%

\end{document}